\documentclass[prl,aps,twocolumn,epsfig,superscriptaddress,showpacs]{revtex4}
\usepackage[dvips]{graphicx}

\usepackage{amsmath}

\begin{document}
\author{Jing-Ling Chen}
 \email{chenjl@nankai.edu.cn}
\affiliation{Theoretical Physics Division, Chern Institute of
Mathematics, Nankai University, Tianjin 300071, People's Republic of
China}
\author{Dong-Ling Deng}
 \affiliation{Theoretical Physics Division, Chern Institute of
Mathematics, Nankai University, Tianjin 300071, People's Republic of
China}
\author{Ming-Guang Hu}
\affiliation{Theoretical Physics Division, Chern Institute of
Mathematics, Nankai University, Tianjin 300071, People's Republic of
China}

\date{\today}

\title{SO(4) symmetry in the relativistic hydrogen atom}

\begin{abstract}
We show that the relativistic hydrogen atom possesses an SO(4)
symmetry by introducing a kind of pseudo-spin vector operator. The
same SO(4) symmetry is still preserved in the relativistic quantum
system in presence of an U(1) monopolar vector potential as well as
a nonabelian vector potential. Lamb shift and SO(4) symmetry
breaking are also discussed.
\end{abstract}



\maketitle

 Symmetry principle is one of the cornerstones of modern physics. It
has been playing a more and more significant role in theoretical
physics since the early twentieth century, when Einstein first put
it as the primary feature of nature that constrains the allowable
dynamical laws \cite{D.J.Gross}. The Einstein's profound change of
attitude on symmetry principle has made a great progress in the
study of symmetry. In the latter half of the twentieth century it
has become the most dominant concept in the exploration and
formulation of the fundamental laws of physics, and nowadays it
serves as a guiding principle in the search for further unification
theory. Dynamical symmetries are prevalent in many important
physical models. For instance, in the nonrelativistic quantum
mechanics, a three-dimensional harmonic oscillator has an U(3)
symmetry, a three-dimensional hydrogen atom has an SO(4) symmetry
\cite{Lange}, and the Haldane-Shastry model, which describes the
one-dimensional long-range spin-interaction chain, has a Yangian
symmetry \cite{Haldane}.

As is well-known that the nonrelativistic quantum mechanics is an
approximate theory of the relativistic one. This gives rise to a
fundamental open question: Suppose a nonrelativistic quantum system
possesses a certain dynamical symmetry, when its corresponding
relativistic quantum mechanical version  is taken into account, will
the same symmetry still reside in the system? Harmonic oscillator
and hydrogen atom are the two simplest prototype models in quantum
physics. Dirac himself has introduced a kind of relativistic version
for the quantum mechanical harmonic oscillator with the Hamiltonian
$H=\vec{\alpha}\cdot (\vec{p}-i \beta M \omega \vec{r})+\beta M$,
which now known as the Dirac oscillator \cite{Lange}. However, so
far the full symmetry of the Dirac oscillator has not yet been
clear. Very recently, Ginocchio has made a remarkable progress
\cite{Joseph N. Ginocchio} by showing that U(3) symmetry does reside
in a kind of relativistic harmonic oscillator, whose Dirac
Hamiltonian reads $H=\vec{\alpha}\cdot \vec{p}+\beta M + (1+\beta) M
\omega^2 r^2/2$, where $\vec{\alpha}$, $\beta$ are the Dirac
matrices, $\vec{p}$ is the three-dimensional linear momentum,
$\vec{r}$ is the spatial coordinate, $r$ its magnitude, $M$ is the
mass, $\omega$ is the frequency, and the velocity of light and the
Planck constant have been set equal to unity, $c=\hbar=1$. This fact
also suggests that Ginocchio's version of the relativistic quantum
harmonic oscillator may be a more natural extension than the Dirac
oscillator.

Whether the relativistic hydrogen atom (RHA) has an SO(4) symmetry
is still open? The purpose of this Brief Report is threefold. First,
we show that there is indeed an SO(4) symmetry in the usual
relativistic hydrogen atom, by introducing a kind of pseudo-spin
vector operator. Second, we illustrate that the same SO(4) symmetry
is still preserved in the relativistic quantum system in presence of
an U(1) monopolar vector potential. Third, we find that the
relativistic hydrogen atom still possesses an SO(4) symmetry if some
kinds of appropriate nonabelian vector potentials are presented.
This reflects that the hydrogen atom (or the Kepler system) is a
highly symmetric system, whatever in the levels of classical
mechanics, quantum mechanics or even the relativistic quantum
theory.

\emph{SO(4) symmetry in the usual RHA.} The Dirac Hamiltonian for a
relativistic hydrogen atom reads
\begin{eqnarray}\label{HRHA1}
H_{rha}=\vec{\alpha}\cdot \vec{p}+\beta M-\frac{a}{r},
\end{eqnarray}
where $\vec{\alpha}=\left(
\begin{matrix}
0 & \vec{\sigma}\\
\vec{\sigma}& 0
\end{matrix}\right)$, $\beta=\left(\begin{matrix}
\bf{1}&0\\
0&-\bf{1}
\end{matrix}\right)$, $\vec{\sigma}$ is the vector of Pauli matrices,
$\bf{1}$ is the $2\times 2$ identity matrix, $a=e^2$ the fine
structure constant, and $e$ the electric charge. Its energy spectra
is given by the Sommerfeld formula \cite{H.Katsura}
\begin{eqnarray}\label{E}
&&\frac{E}{M}=\left(1+\frac{a^2}{(n-|\kappa|+\sqrt{\kappa^2-a^2})^2}\right)^{-1/2},
\\
&&|\kappa|=(j+1/2)=1,2,3,\cdots,\quad n=1,2,3,\cdots\nonumber,
\end{eqnarray}
where $n$ is the radial quantum number, $\kappa=\pm (j+1/2)$ are
eigenvalues of the Dirac's operator
$K=\beta(\vec{\Sigma}\cdot\vec{L}+1)$ with $K^2=\vec{J}^2+1/4$,
$\vec{J}=\vec{L}+\vec{S}$ is the total angular momentum,
$\vec{L}=\vec{r}\times \vec{p}$ the orbital angular momentum, and
$\vec{S}=\vec{\Sigma}/2$ the spin-1/2 angular momentum.
In the nonrelativistic limit, the Hamiltonian (\ref{HRHA1}) reduces
to the usual nonrelativistic hydrogen atom $H_{nrha}=p^2/2M-a/r$,
and the Sommerfeld formula reduces correspondingly to the Bohr
formula $ E_n=-M a^2/2n^2\approx E-Mc^2$.

The Hamiltonian $H_{nrha}$ commutes with $\vec{L}$ and the
well-known Pauli-Runge-Lentz vector \cite{W.Pauli}\cite{M.Bander},
which form a dynamical symmetry group of SO(4). Evidently, the
Bohr's energy formula depends only on the principal quantum number
$n$, and it has $n^2$-fold degeneracies due to the SO(4) symmetry.
However, the Sommerfeld's energy formula is $j$-dependent, for fixed
$n$ and $j$, the energy has only $2(2j+1)$-fold degeneracies for
$n\ne|\kappa|$ and $(2j+1)$-fold degeneracies for $n=|\kappa|$, this
does not support apparently that the RHA still possesses an SO(4)
symmetry, since any reduction or elimination of degeneracy usually
implies the broken of the symmetry. Nevertheless, the Hilbert space
of $H_{rha}$ is larger than that of $H_{nrha}$ by considering the
additional intrinsic spin space. Therefore it is still possible to
restore an SO(4) symmetry for $H_{rha}$ through combining properly
the operators of $\vec{r}$, $\vec{p}$ and $\vec{\Sigma}$.

If this is the case, the question is, what are the six relativistic
generators? What we need to do first is to find out six linear
independent operators (i.e., six integrals of motion for the RHA)
that all commute with $H_{rha}$, and then arrange them to be six
generators of SO(4) group. Up to now, people have known five of
them. The first three are three components of the total angular
momentum operator $\vec{J}$. The fourth is the Dirac's operator $K$
mentioned above. The fifth integral of motion was discovered by
Johnson and Lippmann \cite{M.H.Johnson} in the year of 1950, which
now called the Johnson-Lippmann (JL) operator. Such a famous
discovery has stirred a great furor, and many people have been
attracted in this problem
\cite{L.C.Biedenharn,C.Fronsdal,T.T.Khachidze}. The JL operator
reads \cite{T.T.Khachidze}
\begin{eqnarray}\label{D-K-H}
D=\gamma^5\vec{\alpha}\cdot\hat{r}-\frac{i}{Ma}K\gamma^5(H_{rha}-M\beta),
\end{eqnarray}
with its square is
\begin{equation}\label{A-K-H}
D^2=1+\left(\frac{H_{rha}^2}{M^2}-1\right)\frac{K^2}{a^2},
\end{equation}
where $\gamma^5=\left(\begin{matrix} 0&\bf{1}\\
\bf{1}&0
\end{matrix}\right)$, $\vec{\alpha}=\gamma^5\vec{\Sigma}$, and $\hat{r}=\vec{r}/r$ is a unit vector.
The physical significance of the JL operator in the nonrelativistic
limit is nothing but a projection of the Pauli-Runge-Lentz vector on
the spin angular momentum vector \cite{Lange,T.T.Khachidze}.

The commutation relations among $H_{rha}$ and these five conserved
quantities are $[H_{rha}, \vec{J}]=[H_{rha}, K]=[H_{rha}, D]=0$,
$[\vec{J}, K]=[\vec{J}, D]=0$, and remarkably $\{K, D\}=KD+DK=0$,
namely $K$ and $D$ are anti-commutative. As usual, the simultaneous
eigenfunctions of $\{ H_{rha}, J^2, J_3 \}$ are twofold Krammer's
degeneracies, i.e.,
\begin{eqnarray}\label{E1}
|\psi^+_{njm_j}(\vec{r})\rangle=\frac{1}{\sqrt{\cal
N}}\left(\begin{matrix}
f(r)\phi^A_{jm_j}\\
i g(r)\phi^B_{jm_j}
\end{matrix}\right),\nonumber\\
|\psi^-_{njm_j}(\vec{r})\rangle=\frac{1}{\sqrt{\cal
N}}\left(\begin{matrix}
f(r)\phi^B_{jm_j}\\
i g(r)\phi^A_{jm_j}
\end{matrix}\right),
\end{eqnarray}
with  $H_{rha} |\psi^\pm_{njm_j}\rangle=E |\psi^\pm_{njm_j}\rangle$,
$\vec{J}^2 |\psi^\pm_{njm_j}\rangle=j(j+1)
|\psi^\pm_{njm_j}\rangle$, $J_3 |\psi^\pm_{njm_j}\rangle=m_j
|\psi^\pm_{njm_j}\rangle$, and $m_j$ runs from $-j$ to $j$. Here
${\cal N}=\int_0^{+\infty}dr[f^2(r)+g^2(r)]$ is the normalized
coefficient, $f(r)$ and $g(r)$ are real functions,
$\phi^A_{jm_j}=\frac{1}{\sqrt{2l+1}}\left(\begin{array}{c}
\sqrt{l+m+1}\; Y_{lm}(\theta, \varphi)\\
\sqrt{l-m}\;Y_{l,m+1}(\theta, \varphi)\end{array}\right),
\phi^B_{jm_j}=\frac{1}{\sqrt{2l+3}}\left(\begin{array}{c}
-\sqrt{l-m+1}\;Y_{l+1,m}(\theta, \varphi)\\
\sqrt{l+2+m}\;Y_{l+1,m+1}(\theta, \varphi)\end{array}\right)$,
$Y_{lm}(\theta, \varphi)$ is the spherical harmonics,
and $(\vec{\sigma}\cdot\hat{r})\phi^A_{jm_j}=-\phi^B_{jm_j}$,
$(\vec{\sigma}\cdot\hat{r})\phi^B_{jm_j}=-\phi^A_{jm_j}$. The
eigenstates $|\psi^\pm_{njm_j}(\vec{r})\rangle$ are distinguished by
the Dirac's operator as $K |\psi^\pm_{njm_j}(\vec{r})\rangle= \pm
|\kappa| |\psi^\pm_{njm_j}(\vec{r})\rangle$. The existence of the JL
operator is the direct reason that causing the twofold Krammer's
degeneracies \cite{M.H.Johnson}. Later on we shall show this fact
from the viewpoint of the pseudo-spin operators.

The anti-commutativity between operators $K$ and $D$ motivates us to
introduce the pseudo-spin vector operator $\vec{T}=(T_1, T_2,
T_3)=(\tau_1, \tau_2, \tau_3)/2$, where
\begin{equation}
\tau_1=\frac{D}{\sqrt{D^2}},\quad
\tau_2=\frac{iDK}{\sqrt{D^2K^2}},\quad \tau_3=\frac{K}{\sqrt{K^2}}.
\end{equation}
The operators $\tau_3$ and $\tau_1$ are defined by rescaling $K$ and
$D$ such that $\tau_3^2=\tau_1^2=1$. The operator $\tau_2$ is
defined by the commutator $[\tau_3, \tau_1]=2 i \tau_2$, or
$\tau_2=i\tau_1 \tau_3$. The vector operator $\vec{\tau}$ plays a
similar role as the Pauli matrices vector $\vec{\sigma}$. If
$|\psi^+_{njm_j}(\vec{r})\rangle$ is an eigenstate of $H_{rha}$ and
$\tau_3$, then $\tau_1 |\psi^+_{njm_j}(\vec{r})\rangle$ is also an
eigenstate of $H_{rha}$ and $\tau_3$ because of
$\tau_1\tau_3\tau_1=-\tau_3$. This is the reason causing the twofold
Krammer's degeneracies.

It is easy to show that $[T_i, T_j]=i\epsilon_{ijk} T_k$ and
$T^2=\frac{1}{2}(\frac{1}{2}+1)$. The vector operator $\vec{T}$ has
a property like spin-$1/2$, yet it is not a spin, because it
contains $\vec{r}$ and $\vec{p}$, consequently we call it a
pseudo-spin-1/2 vector operator. One may have $[J_i, T_j]=0$, in
other words, $\vec{J}$ and $\vec{T}$ are two independent angular
momentum vectors that commute with the Hamiltonian $H_{hra}$.
Therefore, after making the following simple linear combinations
\begin{equation}\label{De:LRL}
\vec{I}=\vec{J}+\vec{T}, \;\;\; \vec{R}=\vec{J}-\vec{T},
\end{equation}
one arrives at an SO(4) algebraic relation: $[I_i,
I_j]=i\epsilon_{ijk} I_k$, $[I_i, R_j]=i\epsilon_{ijk} R_k$, $[R_i,
R_j]=i\epsilon_{ijk} I_k$. Since $[H_{rha}, \vec{I}]=[H_{rha},
\vec{R}]=0$, this ends the finding of SO(4) dynamical symmetry in
the usual RHA.

Full energy spectra have been derived by the well-known
ladder-operator procedure \cite{Lange}\cite{H.Katsura}. The symmetry
involved in the system is helpful for us to obtain the energy
spectra. For example, in Eq. (\ref{A-K-H}), since $D^2$ is
positively defined, its minimal eigenvalue is zero, then one can get
precisely the ground state energy of hydrogen atom
\cite{T.T.Khachidze} as
$E|_{n=\kappa=1}=M\left(1-a^2\right)^{1/2}$. Furthermore, with the
aid of the pseudo-spin-1/2 vector operator, one can establish an
elegant formula between energy spectra $E$ and the integrals of
functions $f(r)$ and $g(r)$. More explicitly, let us denote
$|\pm\rangle \equiv|\psi^\pm_{njm_j}\rangle$, which are eigenstates
of $T_3$ with eigenvalues equal to $\pm 1/2$. $T_+=T_1 + iT_2$ is
the raising operator $T_+|-\rangle=|+\rangle$, hence $\langle
+|T_+|-\rangle=1$. It is easy to check that
$T_+|-\rangle=\frac{D}{\sqrt{D^2}}|-\rangle=\frac{1}{\sqrt{D^2}}\frac{1}{\sqrt{\cal
N}} \left(
\begin{array}{c}
-f(r)\phi^A_{jm_j}+\frac{1}{Ma}(E+M)(j+\frac{1}{2})g(r)\phi^A_{jm_j}\\
-\frac{i}{Ma}(E-M)(j+\frac{1}{2})f(r)\phi^B_{jm_j}-ig(r)\phi^B_{jm_j}
\end{array}\right).$
Thus,
$\langle+|T_+|-\rangle
=\frac{-1}{\sqrt{D^2}}\left(1-\frac{j+1/2}{{\cal
N}a}\int_0^{+\infty}dr[2f(r)g(r)]\right) =1$.
Let $\int_0^{+\infty}dr[2f(r)g(r)]/{\cal N}=b$, we then obtain an
equation
$1+\sqrt{D^2}=(j+1/2)b/a$.
By using Eq. (\ref{A-K-H}), we obtain the desired formula
\begin{eqnarray}\label{AI}
\frac{E^2}{M^2}=b^2-\frac{2a}{j+\frac{1}{2}}b+1.
\end{eqnarray}
The exact solutions of $f(r)$ and $g(r)$ are
\cite{Bethe}\cite{Robson}:
\begin{eqnarray}\label{fg}
f(r)&=&\sqrt{M+E}[-\tilde{n}F(1-\tilde{n},2\nu+1,\rho)+(Ma\lambda+\kappa)\times\nonumber\\
&&F(-\tilde{n},2\nu+1,\rho)]\; \rho^{\nu-1}e^{-\rho/2},\nonumber\\
g(r)&=&\sqrt{M-E}[-\tilde{n}F(1-\tilde{n},2\nu+1,\rho)-(Ma\lambda+\kappa)\times\nonumber\\
&&F(-\tilde{n},2\nu+1,\rho)]\; \rho^{\nu-1}e^{-\rho/2},
\end{eqnarray}
where $\tilde{n}=n-|\kappa|$, $\nu=\sqrt{K^2-a^2}$,
$\rho=2r/Ma\lambda$, $\lambda=1/\sqrt{M^2-E^2}$. Substituting Eq.
(\ref{fg}) into Eq. (\ref{AI}), one can verify directly the
correctness of Eq. (\ref{AI}).

\emph{SO(4) symmetry in the RHA with an U(1) monopole.} In order to
find the reason for the existence of the smallest electric charge,
Dirac published a paper in $1931$ which started the subject of
magnetic monopoles \cite{Dirac}. After $1931$, the theory of
monopoles
has been studied extensively in many literatures, such as
\cite{I.Tamm,T.T.Wu}. The Dirac Hamiltonian of the U(1)-monopolar
RHA reads
\begin{eqnarray}\label{De:H_mon}
H'_{rha}=\vec{\alpha}\cdot\vec{\pi}+m\beta-\frac{a}{r},
\end{eqnarray}
where $\vec{\pi}=\vec{p}-e\vec{A}$, $\vec{A}$ is the vector
potential of an U(1) Wu-Yang monopole with strength $g$
\cite{T.T.Wu}, and
$\vec{B}=\nabla\times\vec{A}=g\frac{\vec{r}}{r^3}$ is the magnetic
field satisfying the Coulomb gauge $\nabla \cdot \vec{A}=0$.
Similarly, one finds that $H'_{rha}$ possesses an SO(4) symmetry,
the corresponding generators are $\vec{I}'=\vec{J}'+\vec{T}'$,
$\vec{R}'=\vec{J}'-\vec{T}'$, where $\vec{J}'=\vec{L}'+\vec{S}$,
$\vec{L}'=\vec{r}\times\vec{\pi}-q \hat{r}$ is the
monopole-dependent orbital angular momentum vector,
$q=eg=\frac{1}{2}\times {\rm integer}$ is the magnetic charge,
$\vec{T}'=(D'/\sqrt{{D'}^2}, i D'K'/\sqrt{{D'}^2 {K'}^2},
K'/\sqrt{{K'}^2})/2$,
$K'=\beta[\vec{\Sigma}\cdot(\vec{r}\times\vec{\pi})+1]=\beta[\vec{\Sigma}\cdot\vec{L'}+q
\vec{\Sigma}\cdot\hat{r}+1]$ is the monopole-dependent Dirac's
operator, $K'^2=\vec{J'}^2+1/4-q^2$,
$D'=\gamma^5\vec{\alpha}\cdot\hat{r}-({iK'}/{Ma})\gamma^5(H'_{rha}-M\beta)$
is the monopole-dependent JL operator, and
${D'}^2=1+({H'}_{rha}^2/{M^2}-1)({{K'}^2}/{a^2})$.
It is worthy to mention that $D'$ and ${D'}^2$ keep the same
structures of $D$ and ${D}^2$.

The SO(4) symmetry of ${H'}_{rha}$ can be checked directly.
Correspondingly, its energy spectra is given by the monopolar
Sommerfeld formula
\begin{eqnarray}\label{Emo}
&&\frac{E'}{M}=\left(1+\frac{a^2}{(n'-|\kappa'|+\sqrt{{\kappa'}^2-a^2})^2}\right)^{-1/2},
\end{eqnarray}
where $\kappa'=\pm\sqrt{(j'+1/2+q)(j'+1/2-q)}$  are eigenvalues of
the Dirac's operator $K'$, and $n'=0+|\kappa'|$, $1+|\kappa'|$,
$2+|\kappa'|$, $\cdots$. When $q=0$, all the extended
monopole-dependent operators and relations reduce to the usual ones
in RHA. Furthermore, in the nonrelativistic limit, the Hamiltonian
(\ref{De:H_mon}) reduces to the nonrelativistic monopolar-hydrogen
atom $H'_{nrha}=\vec{\pi}^2/2M+q^2/2Mr^2-a/r$. Such a quantum system
still has an SO(4) symmetry due to the monopole-dependent orbital
angular momentum vector and the monopole-dependent Pauli-Runge-Lentz
vector \cite{J.L.Chen}.

\emph{SO(4) symmetry in the RHA with a nonabelian vector potential.}
Let us add a nonabelian vector potential $\vec{\cal A}=iW(r)
\vec{\Sigma}\times\vec{r}$ to RHA, where $W(r)$ ia an arbitrary real
function of $r$. Then the Dirac Hamiltonian reads
\begin{eqnarray} H^{''}_{rha}
=\vec{\alpha}\cdot (\vec{p}-e\vec{\cal A})+\beta m-\frac{a}{r}.
\end{eqnarray}
Similarly, $H^{''}_{rha}$ possesses an SO(4) symmetry with the
corresponding generators are $\vec{I}^{''}=\vec{J}+\vec{T}^{''}$,
$\vec{R}^{''}=\vec{J}-\vec{T}^{''}$. Here $\vec{J}$ is the usual
total angular momentum operator,
$\vec{T}^{''}=(D^{''}/\sqrt{{D^{''}}^2}, i D^{''}K/\sqrt{{D^{''}}^2
{K}^2}, K/\sqrt{{K}^2})/2$,
$D^{''}=\gamma^5\vec{\alpha}\cdot\hat{r}-({iK}/{Ma})\gamma^5(H^{''}_{rha}-M\beta)$
is the JL operator in presence of the nonabelian vector potential,
and ${D^{''}}^2=1+({{H^{''}}_{rha}^2}/{M^2}-1)({{K}^2}/{a^2})$.
Also the operators $D^{''}$ and ${D^{''}}^2$ share the same forms of
$D$ and ${D}^2$. The nonabelian vector potential satisfies the
Coulomb gauge $\nabla \cdot \vec{\cal A}=0$, from ${\cal
B}_i=(1/2)\epsilon_{ijk}(\partial_j {\cal A}_k-\partial_k {\cal A}_j
+ [{\cal A}_j, {\cal A}_k])$ one has the ``magnetic" field as
$\vec{\cal
B}=i[2W(r)-rW'(r)]\vec{\Sigma}+i[2W^2(r)+W'(r)/r](\vec{\Sigma}\cdot\vec{r})\vec{r}$.
Interestingly, if we choose $W(r)=-1/r^2$, then the ``magnetic"
field $\vec{\cal B}=(-4i/r^2)\vec{\Sigma}$ is proportional to the
spin vector.

\emph{Lamb shift and SO(4) symmetry breaking.}
The spectrum formula (\ref{E}) indicates that the levels having the
same $n$ and $j$ values should be degenerate, for instance, the
$2S_{1/2}$ and $2P_{1/2}$ levels should share the same energy.
However, In $1947$ Lamb and Retherford made an elaborate experiment
and it showed that the $2P_{1/2}$ energy level was depressed about
$1.057\times 10^9$ Hz below the $2S_{1/2}$ energy level
\cite{Lambshift}. This effect is now called the Lamb shift. It
cannot be explained in the framework of the ordinary quantum
mechanics and gives rise to the birth of the quantum electrodynamics
(QED). In fact, the successful calculation of these small quantum
corrections to the Dirac energy levels was one of the remarkable
achievements of quantum field theory. From the viewpoint of QED,
Lamb shift is caused by the vacuum polarization and vertex
corrections \cite{M.Kaku}, and the effective potential reads
\begin{eqnarray}\label{Lamb}
\triangle V_{Lamb}\approx\frac{4a^2}{3M^2}\left(\ln\frac{M}{\mu}
-\frac{1}{5}\right)\delta^3(\vec{r})
+\frac{a^2(\vec{\Sigma}\cdot\vec{L})}{4\pi M^2 r^3}.
\end{eqnarray}
the second term represents the interaction between spin and orbital
angular momentum. Obviously, the corrected Hamiltonian ${\cal
H}_{rha}=H_{rha}+ \triangle V_{Lamb}$ commutes only with $\vec{J}$
and $K$, thus the SO(4) symmetry is broken.

\begin{figure}
\includegraphics[width=83mm]{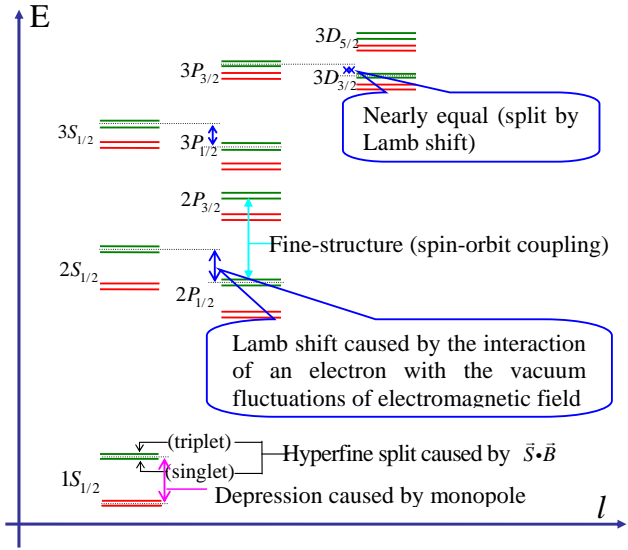}\\
 \caption{(Color online) Low-lying energy levels of $H_{rha}$ and $H'_{rha}$ with $q=1/2$
 (not drawn to scale).
 The green lines denote the energy spectra $E$ in Eq. (\ref{E}), while the red lines denote
 the energy spectra $E'$ in Eq. (\ref{Emo}). The energy levels of $H'_{rha}$  are depressed
below  those of $H_{rha}$,  and such depressions fade away as $n$,
$j$ tend to infinite. For $n=2$, $j=1/2$, the depression caused by
monopole ($\approx10^{14}$ Hz) is much bigger than that of Lamb
shift ($\approx10^9$ Hz). }\label{fig:1}
\end{figure}

In Fig.1, we have plotted some low-lying energy levels of $H_{rha}$
and $H'_{rha}$ with $q=1/2$  (not drawn to scale). The green lines
denote the energy spectra $E$ in Eq. (\ref{E}), while the red lines
denote the energy spectra $E'$ in Eq. (\ref{Emo}).
One finds that $E'_{n'j'}$ are depressed below  $E_{nj}$. For
instance, let $\Delta^q_{nj}=E_{nj}-E'_{n'j'}$ denote the
depressions, then for $q=1/2$ one has the depression for the ground
state as $\Delta^{1/2}_{1,1/2}=1.098\times 10^{15}$ Hz, and that of
the first excited state as $\Delta^{1/2}_{2,1/2}=1.225\times
10^{14}$ Hz. Actually, such depressions fade away as $n$, $j$ become
larger, e.g., the depression of the levels $3D_{5/2}$ ($1.294\times
10^{12}$ Hz)
is only about one thousandth of that of $1S_{1/2}$ ($1.098\times
10^{15}$ Hz).
For fixed $n$, $j$, the depressions caused by monopole are much
bigger than those of Lamb shift.

In conclusion, we have shown that the  relativistic hydrogen atom
possesses an SO(4) symmetry by introducing a kind of pseudo-spin
vector operator. The same SO(4) symmetry is still preserved in the
relativistic quantum system in presence of an U(1) monopolar vector
potential as well as a nonabelian vector potential. When the effect
of Lamb shift is taken into account, the SO(4) symmetry in the
quantum system is broken.


This work is supported in part by NSF of China (Grant No. 10575053
and No. 10605013) and Program for New Century Excellent Talents in
University.

\end{document}